\documentclass[twocolumn,showpacs,amsmath,amssymb,prl,superscriptaddress,floatfix]{revtex4}
\usepackage{graphicx}
\usepackage{dcolumn}
\usepackage{bm}
\usepackage{verbatim}
\newcommand{\dd}{\mbox{\textrm{d}}}

\begin{document}
\title{{Kaon pair production in proton-nucleus collisions at 2.83~GeV kinetic energy}}

\author{Yu.T.~Kiselev}\email[E-mail: ]{yurikis@itep.ru}%
\affiliation{Institute for Theoretical and Experimental Physics,
RU-117218 Moscow, Russia}
\author{M.~Hartmann}\email[E-mail: ]{m.hartmann@fz-juelich.de}%
\affiliation{Institut f\"ur Kernphysik, Forschungszentrum J\"ulich, D-52425 J\"ulich, Germany}
\author{A.~Polyanskiy}%
\affiliation{Institute for Theoretical and Experimental Physics,
RU-117218 Moscow, Russia}
\affiliation{Institut f\"ur Kernphysik, Forschungszentrum J\"ulich, D-52425 J\"ulich, Germany}
\author{E.Ya.~Paryev}
\affiliation{Institute for Theoretical and Experimental Physics,
RU-117218 Moscow, Russia}
\affiliation{Institute for Nuclear Research, Russian Academy of
Sciences, RU-117312 Moscow, Russia}
\author{S.~Barsov}
\affiliation{High Energy Physics Department, Petersburg Nuclear
Physics Institute, RU-188350 Gatchina, Russia}
\author{M.~B\"uscher}
\affiliation{Peter Gr\"unberg Institut (PGI-6), Forschungszentrum J\"ulich, D-52425 J\"ulich, Germany}
\author{S.~Dymov}
\affiliation{Institut f\"ur Kernphysik, Forschungszentrum J\"ulich, D-52425 J\"ulich, Germany}
\affiliation{Laboratory of Nuclear Problems, Joint Institute for
Nuclear Research, RU-141980 Dubna, Russia}
\author{R.~Gebel}
\affiliation{Institut f\"ur Kernphysik, Forschungszentrum J\"ulich, D-52425 J\"ulich, Germany}
\author{V.~Hejny}
\affiliation{Institut f\"ur Kernphysik, Forschungszentrum J\"ulich, D-52425 J\"ulich, Germany}
\author{A.~Kacharava}
\affiliation{Institut f\"ur Kernphysik, Forschungszentrum J\"ulich, D-52425 J\"ulich, Germany}
\author{I.~Keshelashvili}
\affiliation{Institut f\"ur Kernphysik, Forschungszentrum J\"ulich, D-52425 J\"ulich, Germany}
\author{B.~Lorentz}
\affiliation{Institut f\"ur Kernphysik, Forschungszentrum J\"ulich, D-52425 J\"ulich, Germany}
\author{Y.~Maeda}
\affiliation{Research Center for Nuclear Physics, Osaka
University, Ibaraki, Osaka 567-0047, Japan}
\author{S.~Merzliakov}
\affiliation{Institut f\"ur Kernphysik, Forschungszentrum J\"ulich, D-52425 J\"ulich, Germany}
\affiliation{Laboratory of Nuclear Problems, Joint Institute for
Nuclear Research, RU-141980 Dubna, Russia}
\author{S.~Mikirtytchiants}
\affiliation{Institut f\"ur Kernphysik, Forschungszentrum J\"ulich, D-52425 J\"ulich, Germany}
\affiliation{High Energy Physics Department, Petersburg Nuclear
Physics Institute, RU-188350 Gatchina, Russia}
\author{H.~Ohm}
\affiliation{Institut f\"ur Kernphysik, Forschungszentrum J\"ulich, D-52425 J\"ulich, Germany}
\author{V.~Serdyuk}
\affiliation{Institut f\"ur Kernphysik, Forschungszentrum J\"ulich, D-52425 J\"ulich, Germany}
\affiliation{Laboratory of Nuclear Problems, Joint Institute for
Nuclear Research, RU-141980 Dubna, Russia}
\author{A.~Sibirtsev}\thanks{Deceased}
\affiliation{Department of Physics and Astronomy, University of Manitoba, Winnipeg,
Manitoba R3T 2N2, Canada}
\author{V.Y.~Sinitsyna}
\affiliation{P.~N.~Lebedev Physical Institute, RU-119991 Moscow, Russia}
\author{H.J.~Stein}
\affiliation{Institut f\"ur Kernphysik, Forschungszentrum J\"ulich, D-52425 J\"ulich, Germany}
\author{H.~Str\"oher}
\affiliation{Institut f\"ur Kernphysik, Forschungszentrum J\"ulich, D-52425 J\"ulich, Germany}
\author{S.~Trusov}
\affiliation{Institut f\"ur Kern- und Hadronenphysik,
Helmholtz-Zentrum Dresden-Rossendorf, D-01314 Dresden, Germany}
\affiliation{Skobeltsyn Institute of Nuclear Physics, Lomonosov Moscow
State University, RU-119991 Moscow, Russia}
\author{Yu.~Valdau}
\affiliation{High Energy Physics Department, Petersburg Nuclear
Physics Institute, RU-188350 Gatchina, Russia}
\affiliation{Helmholtz-Institut f\"ur Strahlen- und Kernphysik, Universit\"at
Bonn, D-53115 Bonn, Germany}
\author{C.~Wilkin}
\affiliation{Physics and Astronomy Department, UCL, London WC1E 6BT,
United Kingdom}
\author{P.~W\"ustner}
\affiliation{Zentralinstitut f\"ur Elektronik,
Forschungszentrum J\"ulich, D-52425 J\"ulich, Germany}
\author{Q.J.~Ye}
\affiliation{Institut f\"ur Kernphysik, Forschungszentrum J\"ulich, D-52425 J\"ulich, Germany}
\affiliation{Department of Physics and Triangle Universities Nuclear Laboratory,
Duke University, Durham, NC 27708, USA}

\date{\today}

\vspace*{1cm}
%
\begin{abstract}
The production of non-$\phi$ $K^+K^-$ pairs by protons of 2.83~GeV kinetic
energy on C, Cu, Ag, and Au targets has been investigated using the COSY-ANKE
magnetic spectrometer. The $K^-$ momentum dependence of the differential
cross section has been measured for laboratory polar angles
$\theta_{K^{\pm}}\le 12^{\circ}$ over the 0.2--0.9~GeV/$c$ range. The
comparison of the data with detailed model calculations indicates an
attractive $K^-$-nucleus potential of about $-60$~MeV at normal nuclear
matter density at a mean momentum of 0.5~GeV/c. However, this approach has
difficulty in reproducing the smallness of the observed cross sections at low
$K^-$ momenta.
\end{abstract}

\maketitle

%
\section{Introduction}

The study of kaon and antikaon properties in a strongly interacting
environment has been a very active research field over the last two decades
(see, \emph{e.g.}, \cite{Cassing:1999,Friedman:2007,Hartnack:2012}),
especially in connection with questions of the partial restoration of chiral
symmetry in hot or dense nuclear matter and of the existence of a $K^-$
condensate in neutron stars.

It is reasonably well
established~\cite{Cassing:1999,Friedman:2007,Hartnack:2012} that the $K^+$
meson feels a moderately repulsive nuclear potential of about 20--30~MeV at
normal nuclear matter density, $\rho_0=0.16~\text{fm}^{-3}$. In contrast, the
properties of the $K^-$ meson in nuclear matter are still the subject of very
intense debate. This is due to the complicated dynamics of antikaons inside
nuclei, which lead to modifications of their in-medium properties. These
require complex self-consistent coupled-channel calculations, with the
inclusion of complete sets of pseudoscalar meson and baryon octets. Such
calculations, based on chiral
Lagrangians~\cite{Lutz:1998,Bielich:2000,Ramos:2000,Cieply:2001,Ramos:2001,Tolos:2006,Tolos:2008}
or on meson-exchange potentials~\cite{Tolos:2001,Tolos:2002}, predict
relatively shallow low-energy $K^-$-nucleus potentials with central depths of
the order of $-50$ to $-80$~MeV. On the other hand, fits to the $K^-$ atomic
data~\cite{Friedman:1993,Friedman:1994}, in terms of phenomenological
density-dependent optical potentials or relativistic mean-field
calculations~\cite{Friedman:1999}, lead to much stronger potentials with
depths of about $-200$~MeV at density $\rho_0$. This is in line with the
results obtained in one experiment~\cite{Kishimoto:2007,Kishimoto:2009} but
is in conflict with the self-consistent approaches mentioned above. However,
it should be noted that the antikaonic-atom data probe the surface of the
nucleus and thus do not provide strong constraints on the $K^-$-nucleus
potential at normal nuclear matter density.

Motivated by the idea that a very strong antikaon-nucleon potential could
lead to deeply bound kaonic states~\cite{Akaishi:2002,Akaishi:2005}, many
experiments~\cite{Suzuki:2004,Suzuki:2005,Agnello:2005,Agnello:2006,Agnello:2007,Yamazaki:2010,Fabbietti:2013,Epple:2015,Agakishiev:2015,Sato:2008,Hashimoto:2014,Ichikawa:2015}
have been performed to search for them. Some experiments claim positive
signals~\cite{Suzuki:2004,Suzuki:2005,Agnello:2005,Agnello:2006,Agnello:2007,Yamazaki:2010,Ichikawa:2015}
while others do
not~\cite{Fabbietti:2013,Epple:2015,Agakishiev:2015,Sato:2008,Hashimoto:2014}.
The Valencia theory group has argued that at present there is no firm
experimental evidence for either the existence of deeply bound kaonic states
or for a strong antikaon-nucleus
potential~\cite{Oset:2006,Ramos:2007,Magas:2008,Magas:2009}.

Information about in-medium properties of antikaons can be deduced also from
the study of their production in both heavy-ion and proton-nucleus collisions
at incident energies near or below the free nucleon-nucleon threshold
(2.5~GeV). This can be understood within a scenario where a reduction of the
$K^-$ mass inside the nucleus would lead to an enhancement of the $K^-$ yield
in these collisions, due to in-medium shifts of the elementary production
thresholds to lower energies. However, it was
shown~\cite{Cassing:1999,Hartnack:2012} that the existence of a $K^-$
condensate is not compatible with the available heavy-ion data.

The KaoS data~\cite{Scheinast:2006} on the ratio of $K^-$ and $K^+$ inclusive
momentum spectra from reactions $p+A \to K^{\pm}+X$ with $A=$~C and Au at
laboratory angles from 36$^{\circ}$ to 60$^{\circ}$ and beam energy of
2.5~GeV have been analyzed within the BUU transport
model~\cite{Scheinast:2006}. These calculations have shown that the data are
consistent with an in-medium $K^-A$ potential of the order of $-80$~MeV at
normal nuclear density. This is in agreement with an antikaon attraction of
$-110\pm10$~MeV extracted from heavy-ion
data~\cite{Cassing:1997,Sibirtsev:1998}.

There were measurements at the ITEP accelerator of inclusive antikaon
momentum distributions from 0.6 to 1.3~GeV/$c$ at a laboratory angle of
10.5$^{\circ}$ in $p$Be and $p$Cu interactions at 2.25 and 2.4~GeV beam
energies~\cite{Akindinov:2007,Akindinov:2010}. The $K^-$ excitation functions
in these interactions were also determined for a $K^-$ momentum of
1.28~GeV/$c$ at bombarding energies $< 3$~GeV. A reasonable description of
these data was achieved in the framework of a folding model, based on the
target nucleon momentum distribution and on free elementary cross sections,
assuming vacuum $K^+$ and $K^-$ masses~\cite{Akindinov:2007,Akindinov:2010}.
A $K^-$ potential of about $-28$~MeV at density $\rho_0$ at a momentum of
800~MeV/$c$ has been extracted~\cite{Sibirtsev:1999} from data on elastic
$K^-A$ scattering within Glauber theory.

Given the diverse results, one must admit that the situation with regards to
the antikaon-nucleus optical potential is still very unclear. To make
progress in understanding the strength of the $K^-$ interaction in the
nuclear medium, it is necessary to carry out detailed measurements with
tagged low-momentum $K^-$ mesons. These must not stem from $\phi$ decays so
that they bring ``genuine'' information about this strength. Such
measurements were recently performed by the ANKE Collaboration at COSY, where
the production of $K^+K^-$ pairs with invariant masses corresponding to both
the $\phi$ and non-$\phi$ regions was studied in proton collisions with C,
Cu, Ag, and Au targets at an incident beam energy of
2.83~GeV~\cite{Polyanskiy:2011,Hartmann:2012}. These data allowed the
momentum dependence of the $\phi$ nuclear transparency ratio, the in-medium
$\phi$ meson width, and the differential cross section for its production at
forward angles to be determined for these targets over the $\phi$ momentum
range of 0.6--1.6~GeV/$c$~\cite{Polyanskiy:2011,Hartmann:2012}.

An analysis is here presented of the data from the non-$\phi$ region of
invariant masses, where differential cross sections for $K^+K^-$
pair production on the four targets were obtained as functions of the $K^-$
laboratory momentum. Results of this analysis are compared with model
calculations, based on the nuclear spectral function for incoherent primary
proton-nucleon and secondary pion-nucleon $K^+K^-$ creation processes within
different scenarios for the $K^-$ nuclear potential~\cite{Paryev:2015}.

%
\section{Experiment and Results}

The experiment was performed at the Cooler Synchrotron (COSY) of the
Forschungszentrum J\"{u}lich~\cite{Maier:1997} using the ANKE magnetic
spectrometer~\cite{Barsov:2001,Hartmann:2007} that is located at an internal
target station of the storage ring. ANKE contains three dipole magnets; D1
and D3 divert the circulating beam onto the target and back into the COSY
ring, respectively, while D2 is the analyzing magnet. A series of thin and
narrow C, Cu, Ag, and Au targets was inserted in a circulating beam of
2.83~GeV protons in front of the main spectrometer magnet D2. The ANKE
spectrometer has detection systems placed to the right and left of the beam
to register positively and negatively charged ejectiles which, in the case of
non-resonant kaon pair production, are the $K^+$ and $K^-$. Although only
used here for efficiency studies, forward-going charged particles could also
be measured in coincidence.

\begin{figure}[ht]
  \vspace*{+0mm}
  \includegraphics[clip,width=0.88\columnwidth]{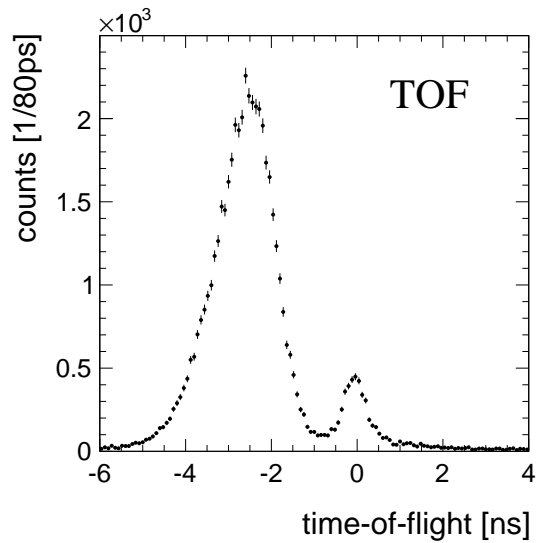}
  \vspace*{-2mm}
  \caption{Time-of-flight difference between the stop counters
    in the negative and positive detection systems for the carbon target.
    The left and right peaks contain the $K^+{\pi}^-$ and $K^+K^-$ events,
    respectively.}
  \label{fig:TOF} \vspace*{-2mm}
\end{figure}

The positively charged kaons were first selected using a dedicated detection
system that can identify a $K^+$ against a pion and/or proton background that
is 10$^5$ times more
intense~\cite{Buescher:2002,Hartmann:2010,Polyanskiy:2012}. The $K^-$ mesons
in correlation with the $K^+$ were subsequently identified from the
time-of-flight difference between stop counters in the negative and positive
detection systems. Figure~\ref{fig:TOF} shows a distribution of such overall
time differences between the negative and positive STOP counters for the
carbon target after correcting for the time delays among different counters,
using information derived from the particle
momenta~\cite{Hartmann:2007,Hartmann:2010,Polyanskiy:2012}.

The peak around zero corresponds to $K^+K^-$ pairs and this sits on a small
background of misidentified particles. The large peak at negative time
differences stems from negative pions, which are faster than the $K^-$ mesons
but are still in coincidence with the $K^+$ mesons registered in the positive
detector. A $3\sigma$ cut around the right peak was made to select the
$K^+K^-$ events. This part of the spectrum is also used to estimate the
residual background for the kaon pairs. The background for the heavier
targets Cu, Ag and Au is noticeably smaller than that for Carbon.

\begin{figure}[h]
  \vspace*{+0mm}
  \includegraphics[clip,width=0.88\columnwidth]{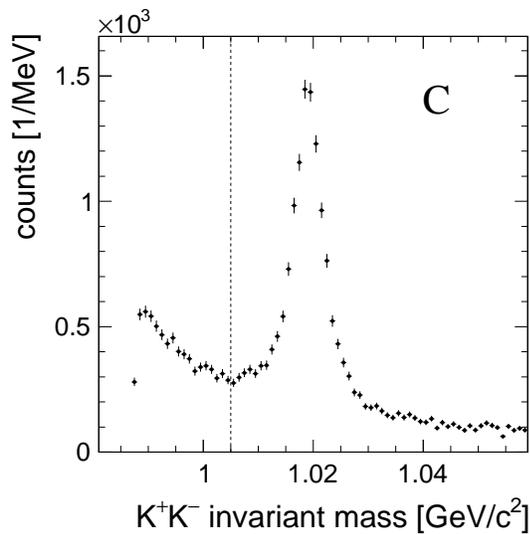}
  \vspace*{-2mm}
\caption{Invariant mass ($\textit{IM}$) distribution for $K^+K^-$ pairs
produced in $p$C collisions at 2.83~GeV beam energy. The vertical line
indicates the cut $\textit{IM} \le 1.005$~GeV/$c$$^2$ used for the separation
of the non-$\phi$ and $\phi$-rich regions.}
  \label{fig:IM} \vspace*{-2mm}
\end{figure}

The resulting invariant mass spectrum of the selected $K^+K^-$ pairs for
Carbon is given in Fig.~\ref{fig:IM}. One can see that there is a strong
$\phi$-signal that sits on a broad distribution of non-$\phi$ kaon pair
production. The invariant mass spectra for Cu, Ag and Au look similar to that
for C~\cite{Polyanskiy:2011,Hartmann:2012,Polyanskiy:2012}. To separate the
non-resonant kaon pair production events from those arising from the decay of
the strong $\phi$ resonance, a cut on the invariant mass of the $K^+K^-$
pairs, $\textit{IM} \le 1.005$~GeV/$c^2$, was applied in the subsequent
analysis. The initial proton kinetic energy of 2.83~GeV corresponds to an
excess energy of 108~MeV above the threshold for kaon pair creation in
proton-nucleon collisions. The accessible ranges of the $K^+$ and $K^-$ meson
momenta were 0.2~GeV/$c$ $ \le p_{K^+} \le 0.6$~GeV/$c$ and 0.2~GeV/$c$ $ \le
p_{K^-} \le 0.9$~GeV/$c$, respectively. The polar production angle was
restricted to 12$^{\circ}$ for both positively and negatively charged kaons.

In order to evaluate the double differential cross section for non-resonant
($\textit{IM} \le 1.005$~GeV/$c^2$) $K^+K^-$ production in $pA$ collisions,
the $K^-$ momentum range was divided into six bins. The numbers
$N_{K^+K^-}^A$ of kaon pairs with the $K^-$ in a momentum bin of width
${\Delta}{p}_{K^-}$ and solid angle ${\Delta}{\Omega}_{K^-}$ in coincidence
with a $K^+$ meson with momentum 0.2~GeV/$c$ $\le p_{K^+} \le$ 0.6~GeV/$c$
and detected in solid angle ${\Delta}{\Omega}_{K^+}$, were determined for the
four targets. The cross section was then evaluated from:
\begin{multline}
\label{eq:CrSecKK}
\frac{\dd^2 \sigma_{pA \to K^+K^-X}} {(\dd p\,\dd\Omega)_{K^+} (\dd p\,\dd\Omega)_{K^-}} = \\
\frac{1} {({\Delta}{p}_{K^+} {\Delta}{\Omega}_{K^+})
({\Delta}{p}_{K^-}{\Delta}{\Omega}_{K^-})} \frac{N_{K^+K^-}^A}
{\langle\epsilon_{K^+K^-}\rangle L_{\rm int}^A},
\end{multline}
where ${\Delta}{p}_{K^+}=0.4$~GeV/$c$,
${\Delta}{\Omega}_{K^{\pm}}=2\pi(1-\cos{12^{\circ}})$ and $L_{\rm int}^A$ is
the integrated luminosity for target $A$.

In order to estimate the average efficiency for $K^+K^-$ identification
$\langle\epsilon_{K^+K^-}\rangle$, the detection efficiency was first
evaluated for each nucleus and each $K^-$ momentum bin. For this purpose the
number of $K^+K^-$ pairs detected relative to that determined from fitting
the $K^+K^-$ efficiency-corrected absolute time-of-flight distributions was calculated
on an event-by-event basis. These efficiencies were then averaged over the
target nuclei for each momentum bin. The root-mean-square deviations of the
individual efficiencies from the $\langle\epsilon_{K^+K^-}\rangle$ mean were
about 5\%, which is consistent with the statistical precision.

\begin{table*}[h!]
\caption{The measured double differential cross sections $\dd^2\sigma_{pA \to
K^+K^-X}/(\dd p\,\dd{\Omega})_{K^+}(\dd p\,\dd{\Omega})_{K^-}$ (in
\mbox{{$\mu$}b}/(GeV/$c$)$^2$sr$^2$) of Eq.~(\ref{eq:CrSecKK}) for
non-resonant $K^+K^-$ production in the interaction of 2.83~GeV protons with
C, Cu, Ag, and Au target nuclei. The data, which are averaged over small kaon
angles, $\theta_{K^{\pm}} \le$ 12$^{\circ}$, and over $K^+$ momenta in the
range $200 \le p_{K^+} \le 600$~MeV/$c$, are presented in bins of $K^-$
momenta. The first errors are statistical and the second systematic, which
are associated with the background subtraction and include the uncertainty in
the average detection efficiency $\epsilon_{K^+K^-}$. There are in addition
overall systematic uncertainties that are discussed in the text. The last
line shows the cross sections (in \mbox{{$\mu$}b}/(GeV/$c$)sr$^2$) integrated
over the total measured $K^-$ momentum range. The related uncertainties are
compounds of the statistical and systematic errors. \label{tab:table1}}
\begin{center}
\renewcommand{\arraystretch}{1.2}
\begin{tabular}{|c|c|c|c|c|}
\hline
 $p_{K^-}$ [MeV/$c$] & C    & Cu             & Ag             & Au   \\
\hline \hline
200--350 &  $2.7\pm0.2\pm0.4$ &  $6.2\pm0.6\pm0.8$ &  $6.5\pm0.8\pm0.7$ & $11.7\pm1.1\pm1.2$ \\
350--450 &  $8.1\pm0.5\pm0.8$ & $18.6\pm1.2\pm1.7$ & $26.2\pm1.8\pm2.4$ & $33.1\pm2.2\pm3.1$ \\
450--550 & $12.7\pm0.6\pm0.8$ & $27.8\pm1.6\pm1.6$ & $32.2\pm2.2\pm1.9$ & $41.7\pm2.8\pm2.4$ \\
550--650 & $11.9\pm0.6\pm0.8$ & $19.3\pm1.4\pm1.3$ & $29.7\pm2.1\pm2.1$ & $30.5\pm2.6\pm2.2$ \\
650--750 &  $6.6\pm0.5\pm0.4$ & $12.3\pm1.2\pm0.8$ & $13.5\pm1.6\pm0.6$ & $18.4\pm2.1\pm2.7$ \\
750--900 &  $2.5\pm0.3\pm0.6$ &  $3.4\pm0.7\pm0.8$ &  $4.4\pm1.0\pm1.2$ &  $6.3\pm1.4\pm1.7$ \\
\hline \hline
200--900 &  $4.7\pm0.2$ &  $9.5\pm0.5$ & $11.8\pm0.6$ & $15.1\pm0.8$ \\
\hline
\end{tabular}
\end{center}
\end{table*}

\begin{figure*}[h]
  \vspace*{+0mm}
  \includegraphics[clip,width=0.8\columnwidth]{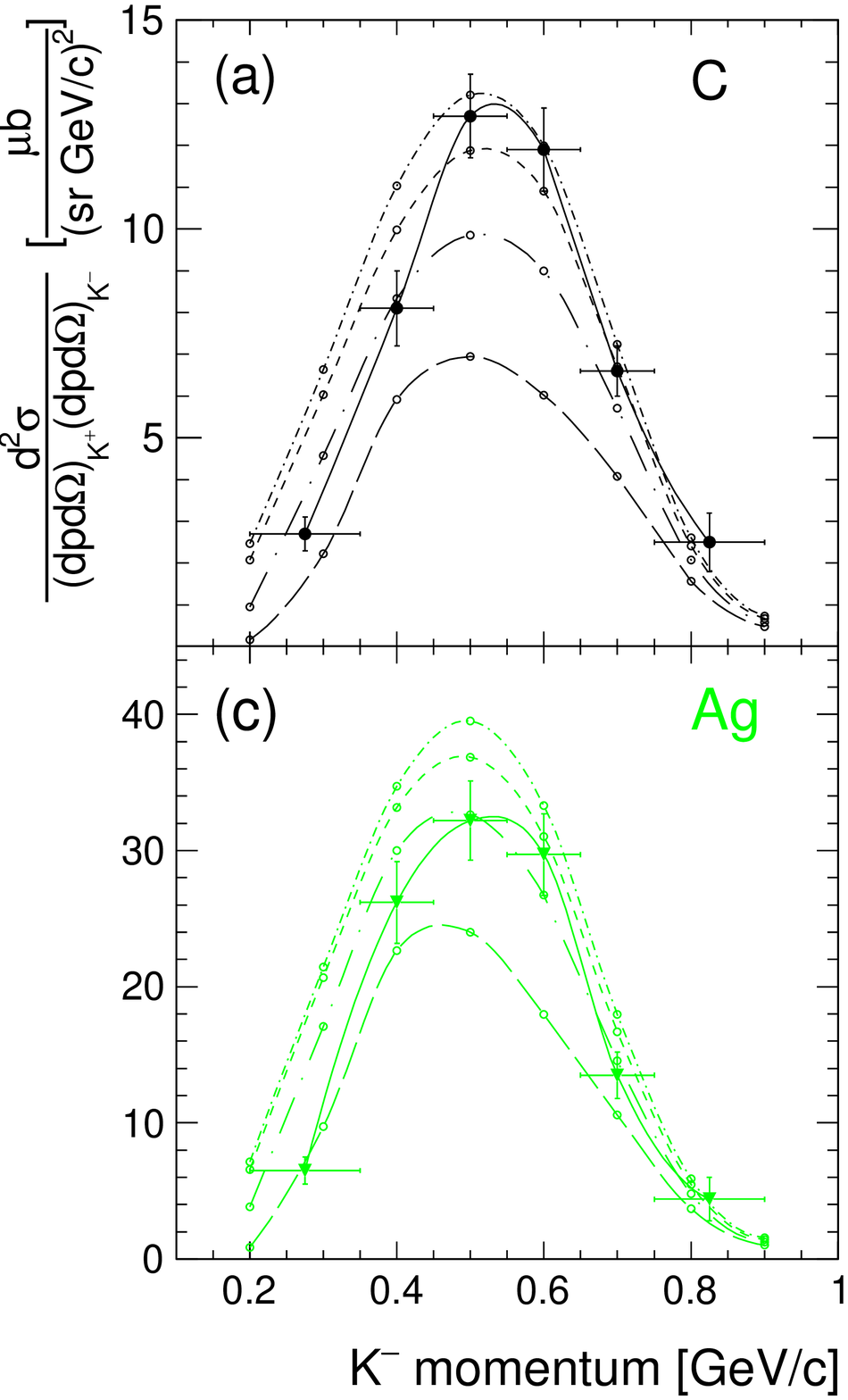}
  \includegraphics[clip,width=0.8\columnwidth]{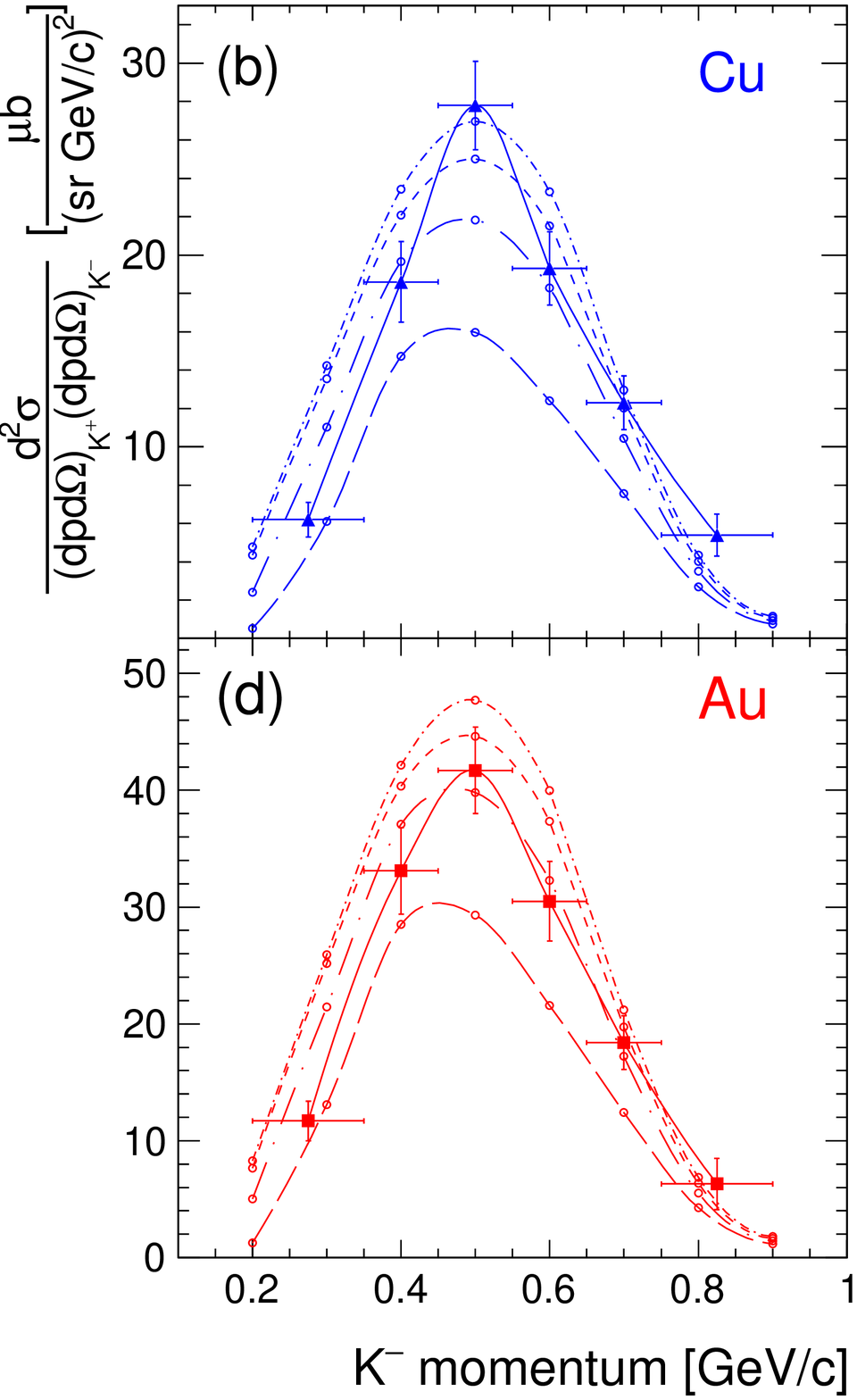}
  \vspace*{-2mm}
\caption{(color online) Double-differential cross sections for the production
of non-resonant $K^+K^-$ pairs in the ANKE acceptance in the collisions of
2.83~GeV protons with C (a), Cu (b), Ag (c), and Au (d) targets as functions
of the $K^-$ laboratory momentum. The experimental data, which are taken from
from Table~\ref{tab:table1}, are averaged over small kaon angles,
$\theta_{K^{\pm}} \le$ 12$^{\circ}$, and over $K^+$ momenta in the range $200
\le p_{K^+} \le 600$~MeV/$c$. The curves represent, from the bottom to top, 
model calculations~\cite{Paryev:2015} for $K^-$ potential depths 
$U=0$~MeV (long-dashed), $-60$~MeV (dot-long dashed), 
$-126$~MeV (short-dashed), and $-180$~MeV (dot-short dashed), respectively.
The solid lines are simple spline functions through the experimental data points.}
  \label{fig:XS}
  \vspace*{-2mm}
\end{figure*}

The overall efficiency was estimated for each event as the product of the
individual efficiencies:
\begin{equation}
\label{eq:EffKK}
\epsilon_{K^+K^-} = \epsilon_{\mathrm{tel}} \times
\epsilon_{\mathrm{tr}} \times
\epsilon_{\mathrm{acc}}.
\end{equation}
The track reconstruction efficiency of $K^+K^-$ pairs $\epsilon_{\rm tr}$ was
determined from the experimental data. The correction for kaon decay in
flight and acceptance, $\epsilon_{\rm acc}$, was estimated as a function of
the laboratory momenta and polar angles of kaons, using simulations. The
range-telescope efficiency $\epsilon_{\rm tel}$ was extracted from
calibration data on $K^+p$ coincidences. The integrated luminosity $L_{\rm
int}^A$ was calculated using the measured flux of $\pi^+$ mesons with momenta
$\approx$ 500~MeV/$c$ produced at small laboratory
angles~\cite{Buescher:2004}.

The statistical uncertainties were about 7\% for each momentum bin and nucleus.
The overall systematic uncertainties were typically 14\%, rising to 16\% for
the first and last momentum bins. The main sources of the systematic effects
are related to the simulation of acceptance corrections $\epsilon_{\rm acc}$ (5\%-10\%),
the determination of the range-telescope efficiency $\epsilon_{\rm tel}$ (10\%), and the
estimation of the integrated luminosity $L_{\rm int}^A$ (8\%).

The measured double-differential cross sections for non-resonant $K^+K^-$ pair
production are given in Table~\ref{tab:table1} for the four targets. The
overall systematic uncertainties of these cross sections have not been included.

%
\section{3. Analysis of Data}

Figure~\ref{fig:XS} shows the measured double-differential cross sections for
$K^+K^-$ production off C, Cu, Ag and Au targets compared to calculations
within the collision model based on the nuclear spectral function for
incoherent primary proton-nucleon and secondary pion-nucleon pair-creation
processes~\cite{Paryev:2015}. The model includes initial proton and final
kaon absorption, using the free $pN$ and $KN$ cross sections, target nucleon
binding and Fermi motion, as well as nuclear mean-field potential effects.
The calculations, which take into account the ANKE acceptance, were performed
assuming four options for the $K^-$ nuclear potential depth $U$ at nuclear
matter density $\rho_0=0.16~\text{fm}^{-3}$, viz.\ $U = 0$~MeV, $U=-60$~MeV,
$U=-126$~MeV, and $U=-180$~MeV.

It is seen from the figure that in general the calculated cross sections for
$K^-$ potential depths $U=-60$, $-126$ and $-180$~MeV follow the data for all
target nuclei for laboratory antikaon momenta above about 0.4~GeV/$c$; the
data exclude the possibility of weak nuclear antikaon mass shifts. The
measured double differential cross sections on light C and medium Cu targets
are better reproduced at these momenta by the model calculations with a
stronger $K^-$ potential. For heavy Ag and Au nuclei the comparison of data
and calculations favors the weaker antikaon potential. On the other hand, the
data at lower antikaon momenta are reproduced reasonably well with almost no
$K^-$ potential and are overestimated by all the calculation with a non-zero
antikaon potential. This suggests that the model misses some peculiarities of
the absorption of low-momentum $K^-$ mesons and/or their production in
nuclear matter.

In the following analysis of the data, aiming at the determination of the
real part of the antikaon nuclear potential at saturation density, we make
use of the cross sections integrated over the measured $K^-$ momentum
interval, \emph{i.e.}, on the last line of Table 1, rather than on the
differential ones shown in Fig.~\ref{fig:XS}. Due to the increased number of
counts, this approach has the advantage of decreasing significantly the
statistical uncertainties to less than about 3\%. In addition, the errors
associated with the background substraction decrease to about 4\%.
This approach also lead to a decrease of the overall systematic uncertainties.
Evidently, the antikaon potential depth extracted in this way will correspond to an
average $K^-$ momentum of about 0.5~GeV/$c$, in the vicinity of which the
main strength of the measured distributions is concentrated. The target mass
dependence of the integrated cross sections follows the power low
$A^{\alpha}$ with a value of exponent $\alpha_{K^+K^-} = 0.42\pm0.02$ which
is less than $\alpha_{\phi} = 0.56\pm0.03$ for the $\phi$
mesons~\cite{Polyanskiy:2011}.

\begin{figure}[hbt]
  \vspace*{+0mm}
  \includegraphics[clip,width=0.88\columnwidth]{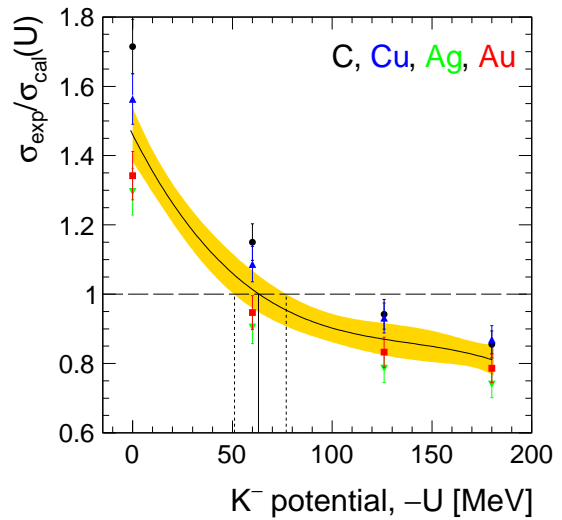}
  \vspace*{-2mm
  }
\caption{(color online) Ratio of the measured integrated cross section for
non-resonant $K^+K^-$ pair production on a given nucleus $A$ to the
corresponding cross sections, calculated within the adopted model supposing
four values for the $K^-$ potential depth at nuclear matter density:
$U=0$~MeV, $-60$~MeV, $-126$~MeV, and $-180$~MeV. The curve represents a
third-order polynomial fit of all ratios presented in the figure, with the
shaded band indicating the 1$\sigma$ confidence interval. The pair of
vertical dotted lines corresponds to the regions where the ratio is unity
within the errors given by the third-order fit. The color code is identical
to that shown explicitly in Fig.~\ref{fig:XS}.}
  \label{fig:POT}
  \vspace*{-2mm}
\end{figure}

To determine the $K^-$ nuclear potential, we consider the ratio of the
measured integrated cross section for the non-resonant $K^+K^-$ pair
production on a given nucleus $A$, as presented in the last line of Table.~1,
to the corresponding cross sections calculated within the model for different
potential strengths. The values of $\sigma_{\rm exp}/\sigma_{\rm cal}(U)$ are
shown in  Fig.~\ref{fig:POT} for $U=0$~MeV, $U=-60$~MeV, $U=-126$~MeV, and
$U=-180$~MeV. Also shown is a third-order polynomial fit to the complete data
set of ratios.

It is seen from the figure that the condition that $\sigma_{\rm
exp}/\sigma_{\rm cal}=1$ is achieved if $U=-(63^{+15}_{-12})$~MeV. However,
this estimate does not include the overall systematic uncertainty in the
data. The calculations have therefore been repeated with the cross sections
increased or decreased by a 13\% uncertainty. This leads to the much expanded
error band of $U=-(63^{+50}_{-31})$~MeV. The width of this band could only be
reduced by controlling better the systematic uncertainties in the values of
the cross sections.

\begin{figure*}[ht]
  \vspace*{+0mm}
  \includegraphics[clip,width=1.6\columnwidth]{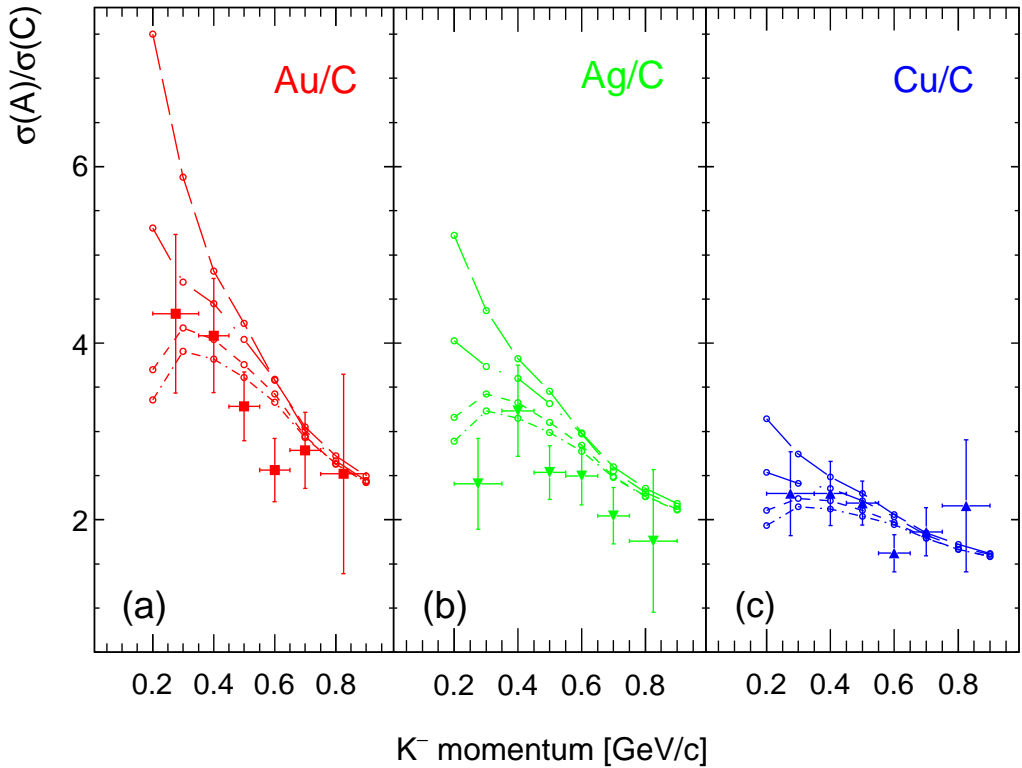}
  \vspace*{-2mm}
\caption{(color online) Ratios of the measured and calculated
double-differential cross sections for non-resonant $K^+K^-$ pair production
off Au (a), Ag (b), and Cu (c) targets presented in Fig.~\ref{fig:XS}, to the measured
and calculated ones for the C target, given also in the same figure as
functions of the $K^-$ laboratory  momentum. The color code and the notation
of the curves are the same as those in Fig.~\ref{fig:XS}.}
  \label{fig:RXS}
  \vspace*{-2mm}
\end{figure*}

Within the uncertainties quoted, the value obtained for the potential depth
is consistent with the moderate $K^-$-nucleus potential of the order of $-50$
to $-80$~MeV that is predicted by calculations based on chiral
Lagrangians~\cite{Lutz:1998,Bielich:2000,Ramos:2000,Cieply:2001,Ramos:2001,Tolos:2006,Tolos:2008}
or on meson-exchange potentials~\cite{Tolos:2001,Tolos:2002}. It also agrees
with the potential of the order of $-80$~MeV at normal nuclear density
extracted from KaoS $pA$ data~\cite{Scheinast:2006}, as well as with a lower
potential of about $-28$~MeV at saturation density extracted at an antikaon
momentum of 800~MeV/$c$~\cite{Sibirtsev:1999}. However, it is hard to
reconcile our value with the deep potential of order $-200$~MeV claimed in
experiments that studied in-flight ($K^-, N$) reactions on $^{12}$C and
$^{16}$O at 1~GeV/$c$~\cite{Kishimoto:2007,Kishimoto:2009}. On the other
hand, it has been argued~\cite{Magas:2010,Ramos:2010} that the ($K^-, N$)
experiment was not suitable for extracting information about the depth of the
$K^-$-nucleus optical potential, though it could provide valuable information
about two and three nucleon absorption mechanisms.

Finally, in Fig.~\ref{fig:RXS} we show the ratios of the measured and
calculated double-differential cross sections for non-resonant $K^+K^-$
production off Cu, Ag, and Au targets to the same for a C target, as
functions of the $K^-$ laboratory momentum. It is worth mentioning that cross
section ratios can be determined with less ambiguity than cross sections
themselves, since the normalization and detector-dependent uncertainties, as
well as theoretical uncertainties associated with the particle production and
absorption mechanisms, largely cancel out. On the other hand, apart from the
sensitivity to the particle absorption in nuclear medium, which is determined
by the imaginary part of particle nuclear potential, such ratios also reveal
some sensitivity to the real part of this potential at low momenta (cf.\
Fig.~\ref{fig:RXS}). The comparison of the strengths and shapes of the data
and calculations provides evidence for a moderately attractive antikaon
optical potential for all the $K^-$ momenta studied. This is in line with our
findings based on the analysis of the integrated cross sections. However, due
to the large errors in the ratios shown in Fig.~\ref{fig:RXS}, these data do
not allow one to get definitive information about the value of this
potential.

%
\section{4. Conclusions}

We have measured the differential cross sections for non-resonant $K^+K^-$
pair production on carbon, copper, silver and gold targets by 2.83~GeV
protons with the ANKE magnetic spectrometer over the antikaon momentum range
of 0.2--0.9~GeV/$c$. In order to determine the $K^-$ nuclear optical
potential we have used a sample of data that is essentially free from
contributions from the strong $\phi$ meson resonance. Information on the
depth of the antikaon nuclear potential was obtained by comparing the
measured cross sections of the non-resonant $K^+K^-$ pair production with
calculations in the framework of a collision model that takes the ANKE
acceptance of Eq.~(\ref{eq:CrSecKK}) into account. It is based on the nuclear
spectral function for incoherent primary proton-nucleon and secondary
pion-nucleon creation processes. Within the model used, the real part of the
attractive $K^-$ nuclear optical potential was found to be about $-60$~MeV at
normal nuclear density and mean $K^-$ momentum of 0.5~GeV/$c$. Although the
error bars are significant, it does not favor a very deep antikaon potential
at this momentum. Further theoretical efforts are needed to reliably describe
the present ANKE data and, hence, to fully elucidate the antikaon dynamics in
the nuclear matter, especially, at low momenta.
\\

%
\begin{acknowledgments}
Support from other members of the ANKE Collaboration, as well as the COSY
machine crew, are gratefully acknowledged. We are particularly appreciative
of the encouragement that we received from Eulogio Oset. This work has been
partially financed by the COSY FFE, the DFG, and the Ministry of Education
and Science of the Russian Federation.
\end{acknowledgments}
%

\end{document}